\documentclass[reprint,aps,pre,superscriptaddress,showkeys,10pt,dcolumn]{revtex4-2}

\usepackage{bm}
\usepackage{lipsum} 
\usepackage{xcolor}
\usepackage{amsbsy}
\usepackage{dcolumn}
\usepackage{tikz}
\usepackage{float}
\usepackage{footnote}
\usepackage{soul}
\usepackage{eqnarray, amsmath, mathrsfs,multirow}

\usepackage{subfigure}
\usepackage{colortbl}
\usepackage{xcolor}
\usepackage{footnote}
\usepackage[flushleft]{threeparttable}
\usepackage{graphicx}
\usepackage{dcolumn}
\usepackage{bm}
\usepackage[colorlinks, linkcolor=black,citecolor=blue,urlcolor=blue]{hyperref}

\begin{document}

\title{Temporal epistasis inference from more than 3,500,000 SARS-CoV-2 Genomic Sequences}

\author{Hong-Li Zeng}
\email{hlzeng@njupt.edu.cn}
\author{Yue Liu}
 \affiliation{School of Science, Nanjing University of Posts and Telecommunications, New Energy Technology Engineering Laboratory of Jiangsu Province, Nanjing, 210023, China}
\author{Vito Dichio}%
\affiliation{%
 Inria Paris, Aramis Project Team, Paris, France\\
 Institut du Cerveau, ICM, Inserm U 1127, CNRS UMR 7225, Sorbonne Université, Paris, France
}%
\author{Erik Aurell}  
\email{eaurell@kth.se}
\affiliation{
 Department of Computational Science and Technology,
AlbaNova University Center, SE-106 91 Stockholm, Sweden
}

\date{\today}

\begin{abstract}
We use Direct Coupling Analysis (DCA) to determine epistatic interactions
between loci of variability of the SARS-CoV-2 virus, segmenting 
genomes by month of sampling. We use full-length, high-quality genomes from the GISAID repository up to October 2021, in total over 3,500,000 genomes. 
We find that DCA terms are more stable over time than correlations, but
nevertheless change over time as mutations disappear from the global population or reach fixation. Correlations 
are enriched for phylogenetic effects, and in particularly 
statistical dependencies at short genomic distances, while DCA
brings out links at longer genomic distance.
We discuss the validity of a DCA analysis under these conditions 
in terms of a transient Quasi-Linkage Equilibrium state.
We identify putative epistatic interaction mutations involving loci in Spike.
\begin{description}
\item[keywords]
SARS-CoV-2 $|$ Temporal Epistasis Inference $|$ Genomic Data $|$ Direct Coupling Analysis
\end{description}
\end{abstract}
                        
\maketitle

\section{\label{sec:level1}Introduction\protect}

The global pandemic of the disease COVID-19
caused by coronavirus SARS-CoV-2
has led to more than 525 million confirmed cases and more than 6 million deaths~\cite{WHO}. 
Efforts to counter the epidemic have include extensive use 
of Non-Pharmaceutical Interventions (NPI)~\cite{Flaxman2020,Salje2020,Kraemer2020,Wong2020,PERRA2021}, and the development of several
vaccines~\cite{TungThanLe-2020-2,AMANAT2020}. To date more than 11,8 billion vaccine doses have been administered
world-wide~\cite{WHO}.
Although drugs such as dexamethasone that lower the fatality rate are now in wide use,
effective anti-viral drugs which would open an another frontline in 
the pandemic are so far lacking~\cite{Gordon2020,Frediansyah2021}.

Both vaccines and antiviral drugs are based on an understanding of the
biology of the pathogen, its strengths and potential weaknesses~\cite{Tse2020,Yoshimoto2020,Hartenian2020}.
The COVID pandemic is the first to have occurred after massive DNA sequencing
became a commodity service. The number of SARS-CoV-2 genomes publicly
available in data repositories is many orders of magnitude larger 
than ever seen in the past. While disparities in sampling and other
sources of bias are serious issues~\cite{Martin2021}, such large amounts
of data should nevertheless be marshaled in support
of the common good to the fullest extent possible. 
In this work we have relied on a full-length high-quality SARS-CoV-2 genome 
sequences from the GISAID repository~\cite{GISAID} with sampling
date up to October 2021: in all more than three and half million viral genomes.
These are the virtually exact genetic blueprints of actual viruses 
infecting actual persons in more than one percent of the confirmed
cases world-wide. That such quasi-real-time monitoring is
at all possible is a staggering achievement.
We are likely only the beginning of the process of understanding
what information that can be unlocked from such vast yet extremely rich and precise data~\cite{Phelan2020}.

Among the most remarkable features of such datasets is the possibility to observe in real time the evolutionary process acting at a population level. In classical population genetics, evolution is driven by four main forces in many aspects analogous to mechanisms of statistical physics \cite{Blythe2007,NeherShraiman2009}.
Mutation is the change of a single genome due to a chance event and can be assimilated to thermal noise.
Natural selection is the propensity of more fit individuals to have more offspring,
and acts as an energy term. 
Recombination (sex) leads to offspring shared between two individuals and acts similarly to pair-wise collisions: the earlier genomes are substituted by partly random new combinations and the distribution over genomes relaxes.
Lastly, genetic drift represents an element of chance, due to the fitness of the population size.

The focus on this work is on epistasis, synergistic 
or antagonistic contributions to fitness from allele variations at two or more loci.
Long-standing theoretical arguments predict that 
the distribution of genotypes in a population
directly reflect such 
multi-loci fitness terms when recombination is the dominant
force of evolution~\cite{Kimura1965,NeherShraiman2011,Zeng2020}.
Coronaviruses in general exhibit recombination due to 
their mode of RNA replication~\cite{LaiCavanagh1997,Graham2010,Robson2020,Gribble2021}, and recombination has
been observed between different strains of SARS-CoV-2
co-infecting the same human host~\cite{Avanzato2020,Baang2021,Choi2020,Hensley2021,Kemp2021}.
While partly conflicting reports have appeared in the literature
as to the impact of recombination on the total SARS-CoV-2 population~\cite{Jackson2021cell,VanInsberghe2021}, it is also the 
case that recent theoretical advances have shown 
similar correspondences also
when mutation is the dominant force
of evolution, provided some recombination is present~\cite{Zeng-Mauri-2021}.
If both mutation and recombination are slower (weaker) processes
than selection, there is on the other hand no simple relation between epistatic contributions
to fitness and variability in the population~\cite{NeherShraiman2009,Neher2013,Dichio2021} and the approach taken
here does not apply.
In this work we assume that the latter scenario does not pertain.

In an earlier contribution we inferred epistatic interactions
from about 50,000 SARS-CoV-2 genome sequences deposited in GISAID
until August 2020~\cite{Zeng2020pnas}. A slightly later 
contribution used about 130,000 sequences available until October 19, 2020,
and reached largely consistent results~\cite{CRESSWELLCLAY2021}.
An important aspect 
of both analyses was to separate
linkage disequilibrium (LD) due to epistasis (the objective of the studies)
and LD due to phylogeny (a confounder).
In~\cite{Zeng2020pnas} interactions imputed to phylogeny were
separated out by a randomized null model procedure~\cite{RodriguezWeigt-2021},
while~\cite{CRESSWELLCLAY2021} leveraged GISAID metadata (sample
geographic position) and assignment of samples to clades.
In this work we used almost two orders of magnitude more data.
This necessitated a different approach, as will be described below. 
Signs of epistasis in large-scale SARS-CoV-2 data was also recently 
investigated by other methods in~\cite{Rochman2021,Rochman2021b}, the authors of
which found limited amounts at the RBD surface of Spike.
This is consistent with the results in~\cite{Zeng2020pnas,CRESSWELLCLAY2021}
where epistasis was mostly detected between loci outside Spike.

In this work we stratify genome sequences as to sampling date.
We collect all sequences sampled in the same month since the beginning
of the pandemic, and analyze epistasis month-by-month.
In contrast to the earlier analysis we find in the new larger data several mutations in Spike
that appear epistatically linked to other mutations in Spike and outside Spike.
Among the highest-ranked such predictions we single out
S:S112L ($21897$), recently associated to vaccine breakthrough infections~\cite{Duerr2021}.
Signs of epistasis in 
data from the wider family of coronaviruses was 
further recently 
investigated in~\cite{Rodriguez-Rivas2022}.
We comment on this important recent and related
contribution in Discussion.

\section{Materials}
\subsection{Data collection}\label{ss-DC}
The input data for this analysis are the genomic sequences of SARS-CoV-2 (high quality and full lengths) as stored in the GISAID~\cite{GISAID} public repository. Each of them is a sequence of $\sim 30, 000$ base pairs (bps), representing either a nucleotide \texttt{A,C,G,T}, or an unknown nucleotide \texttt{N} and a small number of other IUPAC symbols \texttt{KYF} etc., which we will refer to as ``minorities", representing different sets of the aforementioned nucleotides. Any site in the genomic sequence is termed as locus.

The sequences were sorted by collection date - the typical delay with respect to their appearance on GISAID being $> 2$ weeks~\cite{kalia2021lag} - and grouped on a monthly basis until the end of October $2021$.
Considering the small number of sequences available for the first months after the outbreak in the $2020$, data until the end of March $2020$ are grouped together. In total, we hence have $20$ datasets and 3,532,252 sequences.
The number of collected sequences per month is shown in Fig. \ref{fig:seq_num}: it increments towards the $2021$, slightly decreases in the first half on the $2021$, increases again starting from July/August $2021$ and decreases again soon after in September/October.

\subsection{Multiple-Sequence Alignment (MSA)} 
Multiple Sequence Alignments were constructed exploiting the help of the online tool \texttt{MAFFT}~\cite{Katoh2017,Kuraku2013}. Groups of sequences relative to each month are aligned separately with respect to the reference sequence ``Wuhan-Hu-1" - GenBank accession number NC-045512~\cite{Chen2020}.
Note that this is different from what previously done in~\cite{Zeng2020pnas}, where a pre-aligned MSA was used to lighten the computational burden. The resulting MSAs are given as Supplementary Information (SI) Dataset S2, and are also available on the Github repository~\cite{Zeng-github}. 
Each MSA is a matrix $\boldsymbol{\sigma}=\{\sigma_i^n|i=1,...,L, n = 1,...,N_{seqs}\}$, where $N_{seqs}$ represents the number of genomic sequences and varies from month to month by construction, see Fig.~\ref{fig:seq_num}. $L$ columns stand for the genomic loci/sites~\cite{Cocco2018, Horta2019}. Here $L=29,903$ is the total number of loci of the reference sequence. The sites between 256 and 29674 are referred as coding region while others are in the non-coding region. 
Each entry $\sigma_i^n$ of the MSA $\boldsymbol{\sigma}$ is one of the base pairs mentioned in Sec.(\ref{ss-DC}) or a new symbol ``\texttt{-}" introduced to account for a nucleotide deletion or insertion in the alignment process.

\subsection{MSA filtering}
\label{ss-MSAf}
For the MSA filtering, we follow the methods already employed in~\cite{Zeng2020pnas}. As a first step, the ambiguous minorities like \texttt{KYF} etc. are converted into \texttt{N}, so that there are 6 states \texttt{-,N,A,C,G,T} left, which we represent as 0,1,2,3,4,5 respectively. 
Subsequently, all the $20$ MSAs are filtered. In particular, each locus (column) in each MSA is discarded if one of the following two condition is matched: (1) if one same nucleotide is found with a frequency greater than a given $p$ (lack of variability); (2) if the sum of the frequencies of \texttt{A, C, G, T} at this position is less than $0.2$ (non-significant). In Fig.~\ref{fig:num_of_reamin_loci_with_ph_98} the number of survived loci $L_s$ normalized by $N_{seqs}$ for each MSA is shown with $p=0.98$. Similar results with $p=0.9$ and $p=0.999$ are presented  in Fig. \ref{fig:num_of_reamin_loci_with_ph_90} Appendix \ref{app:num_remain_loci} .

\section{Methods}
\subsection{Static Quasi-Linkage-Equilibrium (QLE) phase}\label{subsec:static_QLE}
The QLE state for a genomic population was found by M. Kimura in a study of the steady-state distribution over two bi-allelic loci evolving under selection, mutation and recombination in presence of both additive and epistatic contributions to fitness~\cite{Kimura1965}.
A global QLE state over many loci was reviewed and investigated in~\cite{NeherShraiman2011}, formulating mathematically the evolutionary process in a master-equation 
\begin{equation}\label{e-ME}
    \dot{P}(\bm{\sigma}) = \mathscr{F}_{ev}(\bm{\sigma})\ ,
\end{equation}
where $P$ is a probability distribution in the space of all possible genomic sequences $\bm{\sigma}=\{\sigma_i\}_{i=1}^L$ with $\sigma_i = -1,1$ and $\mathscr{F}_{ev}$ encodes the evolutionary process. This latter approach was further generalized to the case of more than two alleles per locus was considered in~\cite{Gao2019}: here, $\sigma_i = 0,1,\dots,q$ and the fitness function up to pairwise terms reads
\begin{equation}
    F(\bm{\sigma}) = F_0 + \sum_i F_i(\sigma_i) + \sum_{i,j} F_{ij}(\sigma_i,\sigma_j)
\end{equation}

In a static QLE state, the covariance of alleles at each pair of loci is a small but non-zero quantity. In presence of pairwise epistasis $F_{ij}\ne0$ and sufficiently high rate of recombination, the probability distribution $P(\bm{\sigma})$, as it appears in the evolutionary master-equation, reaches a  steady-state distribution takes the Gibbs-Boltzmann form: 
\begin{equation}
 P(\sigma_1,...,\sigma_L)=\frac{1}{Z}e^{-H(\sigma_1,...,\sigma_L)},
 \label{BoltzmannEQ}
\end{equation}
with 
\begin{equation}
 H(\sigma_1,...,\sigma_L) = \sum_{i}h_i(\sigma_i) + \sum_{ij} J_{ij}(\sigma_i,\sigma_j)\ .
 \label{eqn:Ising-Potts}
\end{equation}
In a QLE state, there is a direct relation between the parameters 
describing statistical dependencies in the distribution over genotypes in the population,
and epistasis between loci $i$ and $j$ which give rise to these dependencies, \emph{i.e.,} $J_{ij}(\sigma_i,\sigma_j)\propto F_{ij}(\sigma_i,\sigma_j) $.
The relation has been derived for both bi-allelic and multi-allelic loci,
and if either recombination or mutation is the fastest process~\cite{NeherShraiman2011,Gao2019,Zeng-Mauri-2021}
and have been verified \emph{in silico}~\cite{Zeng2020,Zeng-Mauri-2021} (in simulations).
The relation between correlations and epistasis is on the other hand indirect as it goes through the
relation between correlations and parameters $J_{ij}(\sigma_i,\sigma_j)$ in probability distributions
of this type; a problem variously called ``parameter inference in models in exponential families"~\cite{WainwrightJordan2008},
or ``Direct Coupling Analysis"~\cite{Weigt2009pnas} or ``inverse Ising problem"~\cite{Aurell-2012a,Nguyen2017}. 
In this work our focus is on the parameters $J_{ij}(\sigma_i,\sigma_j)$ themselves.

\subsection{Transient Quasi-Linkage-Equilibrium (QLE) phase}
\label{subsec:transient_QLE}
A QLE state can prevail over a finite time
in the sense that correlations and DCA terms
$J_{ij}$ in \eqref{eqn:Ising-Potts} 
inferred from a temporal snapshot of the population remain stable,
while single-locus frequencies change. 
The mechanism behind such an effect is time-constant
epistatic fitness parameters ($F_{ij}$) coexisting with genetic drift
and/or
time-changing additive fitness parameters ($F_i$).
One scenario when this occurs is when two weakly advantageous
mutations at two different sites appear at about the same time in a population,
and then grow in frequency towards fixation. 
At the very beginning there is only one mutation present, and there is no variability on which epistatic effects can act. 
When both mutations are present but one is still at low
prevalence, both correlation and DCA analysis will give non-zero 
but noisy output due to small sample size.

The equations satisfied by single-locus frequencies and
two-locus joint frequencies in a finite population
were derived in~\cite{NeherShraiman2011} (Eqs.~36 and~37) starting from the same equation (\ref{e-ME}) as above and
under a diffusion approximation and
for an Ising genome model (two alleles per locus).
This approximation is valid when both allele frequencies
at both loci are significant, i.e. none is close to zero
or to one (fixation).
The equations take the form
\begin{eqnarray}
\label{eq:NS-stochastic-i}
\dot{\chi}_i(t) &=& 
(1-\chi_{i}^2) F_i + \sum_{j\neq i} \chi_{ij} F_j - 2\mu \chi_i + \dot{\zeta}_i \\
\label{eq:NS-stochastic-ij}
\dot{\chi}_{ij}(t) &=& \left[(1-\chi_i^2)(1-\chi_j^2) F_{ij} - rc_{ij}\right]\chi_{ij} 
+ \dot{\zeta}_{ij}
\end{eqnarray}
where $\chi_i=\left<\sigma_i\right>$
and $\chi_{ij}=\left<\sigma_i\sigma_j\right>-\chi_i\chi_j$
are signed frequencies and correlations in physical notation,
$F_i$ and $F_{ij}$ are additive and epistatic fitness parameters,
$\mu$ is mutation rate, $r$ overall recombination rate,
$c_{ij}$ is a measure of closeness of loci $i$ and $j$ 
and $\dot{\zeta}_i$ and $\dot{\zeta}_{ij}$ genetic drift noise terms.

It is readily seen that \eqref{eq:NS-stochastic-i} and
\eqref{eq:NS-stochastic-ij} are qualitatively 
different. The first equation describes a process
driven by noise and $(1-\chi_{i}^2) F_i - 2\mu \chi_i$, modulated,
if there are non-zero correlations in the population, by
$\sum_{j\neq i} \chi_{ij} F_j$. 
Depending on
the sign of the net drift, it will hence tend to drive $\chi_i$
towards $\pm 1$ (fixation or elimination of the mutation). 
The second equation on the other hand 
has vanishing drift whenever the expression in the bracket vanishes.
It can be checked that with the small field assumptions
used in~\cite{NeherShraiman2011} when deriving 
\eqref{eq:NS-stochastic-i} and
\eqref{eq:NS-stochastic-ij}, and stated in terms of the
(time-dependent) Ising model parameters, this vanishing of the
bracket corresponds to $J_{ij}=F_{ij}/rc_{ij}$,
and that this is a stable equilibrium
(\cite{NeherShraiman2011}, Eq.~25).
There can thus be a transient QLE phase where single-locus frequencies
may go up in a fluctuating manner for a fairly long time, while $J_{ij}$ and two-mode frequencies
remain steady because governed by a relaxation dynamics.
An extension of the above to the case where the fastest process is mutations
and not recombination can be found in~\cite{Zeng-Mauri-2021}.

Towards the end when one (or both) mutation are close to
fixation both correlation and DCA analysis will give non-zero 
but noisy output due to small sample size,
and \eqref{eq:NS-stochastic-i} and
\eqref{eq:NS-stochastic-ij} are no longer valid.
At the very end when there is only one mutation
left, there is again no variability on which epistatic effects to
act and correlation or DCA analysis applied to the data will 
again yield nothing.

\subsection{Correlation Analysis an LD}
We computed correlations as a measure of
linkage disequilibrium (LD), \textit{i.e.}, as a measure of non-random association between different alleles at different loci. 
For multi-allele distributions, statistical co-variance matrices are defined as 
\begin{equation}
 C_{ij}(a,b)=\left<\mathbf{1}_{\sigma_i,a}\mathbf{1}_{\sigma_j,b}\right>-\left<\mathbf{1}_{
\sigma_i,a}\right>\left<\mathbf{1}_{\sigma_j,b}\right>
 \label{LD}
\end{equation}
where $\mathbf{1}_{\sigma_i,a}=1$ if $\sigma_i=a$ and zero otherwise,
and where $\left<\cdot\right>$ indicates the average over $q$ different
alleles per locus. As discussed above, in our representation of the GISAID data, $q=6$.
We compute overall correlation between site $i$ and $j$ as Frobenius norms
of the statistical co-variance matrices (summation over the inner indices $a,b$ )
\begin{equation}
 C_{ij}=\sqrt{\sum_{a=1}^q \sum_{b=1}^q C_{ij}^2(a,b)}.
 \label{eqn:Frobenius_norm}
\end{equation}

\subsection{plmDCA inference for epistasis between loci}
Correlations differ from statistical dependency encoded in the $J_{ij}$ through \eqref{BoltzmannEQ} - \eqref{eqn:Ising-Potts} because two loci $i$ and $j$ may be correlated even if their direct interaction $J_{ij}$ is zero, provided they both interact with a third locus $k$. Many techniques have been developed to infer the direct couplings in Eq. (\ref{BoltzmannEQ}), see~\cite{Nguyen2017} and references therein. In this work we have used the Pseudo-Likelihood Maximization (plmDCA) 
method~\cite{Besag-1975a,Ravikumar-2010a,Aurell-2012a,Ekeberg-2013a, Ekeberg-2014a, Gao2019}
to estimate the parameters $J_{ij}$.
The basic idea of plmDCA is to substitute maximum-likelihood 
inference of parameters from the joint distribution \eqref{BoltzmannEQ} by the simpler one of estimating which parameters best match the
conditional probabilities
\begin{equation}
 P(\sigma_i|\bm{\sigma}_{\backslash i})=\frac{\exp\left(h_i(\sigma_i) +\sum_{j\ne i}J_{ij}(\sigma_i, \sigma_j)\right)}{\sum_{\mathbf{q}}\exp\left(h_i(q) +\sum_{j\ne i}J_{ij}(q, \sigma_j)\right)};
 \label{conditional_prob}
\end{equation}
Here $\mathbf{q}=\{0,1,2,3,4,5\}$ are the possible states of $\sigma_i$ in the dataset and $\bm{\sigma}_{\backslash i}$ stands for all the loci except the locus $i$. 
Assuming independent samples, the functions to optimize (one for each locus) are
\begin{equation}
 \begin{split}
 \mathscr{P}&\mathscr{L}_i\left(h_i,\{J_{ij}\}_j\right)
 =\\ & \frac{1}{N_{seqs}}\sum_s h_i\left(\sigma_i^{(s)}\right) + \frac{1}{N_{seqs}}\sum_s \sum_{j \ne i} J_{ij}(\sigma_i^{(s)},\sigma_j^{(s)})\\
 &- \frac{1}{N_{seqs}}\sum_s \log\sum_{\mathbf{q}}\exp\left(h_i(q) +\sum_{j\ne i}J_{ij}(q,\sigma_j^{(s)})\right),
\end{split}
\end{equation}
where $s$ labels the sequences (samples), from $1$ to $N_{seqs}$.
We use the asymmetric version of plmDCA~\cite{Ekeberg-2014a} as implemented in~\cite{Gao-github}  with $l_2
$ regularization with penalty parameter $\lambda=0.1$. 
The inferred epistatic interaction between loci $i$ and $j$ is scored by the Frobenius norm 
over the inner indices $a,b$ as in \eqref{eqn:Frobenius_norm}, and as 
implemented in~\cite{Gao-github}. 

\subsection{Removal of phylogenetic confounders}
Statistical dependency between allele distributions at two loci
can arise both from epistatic contributions in QLE, and from inheritance, for example when two unrelated mutations appeared by chance at the same time in a very fit individual spread in the same geographic area (phylogenetic effect).
The global distribution of genotypes then does not have to be of
the Boltzmann form \eqref{BoltzmannEQ}, but can instead reflect mixtures of clones~\cite{Neher2013}.

All data from which one wishes to infer epistasis from LD to some extent 
contain such a combination of the intrinsically epistatic effect, and of phylogeny.
In particular, when recombination acts approximately in the same manner along a genome,
LD due to phylogeny dominates between pairs of loci that are close,
while LD due to epistasis can dominate between pairs of loci that are distant.
In earlier studies on whole-genome data from bacterial pathogens, 
a distance cut-off was therefore employed~\cite{Skwark-2017a,Schubert2019}
as well as in previous work on SARS-CoV-2 data~\cite{Zeng2020pnas,CRESSWELLCLAY2021}.
The effect of phylogenetic correlations in DCA-based contact prediction in proteins
was recently investigated in~\cite{RodriguezWeigt-2021}. 

In the current work we have leveraged the well-documented growth of large clones
in the global SARS-CoV-2 population. 
In particular, we have ascribed large scores $J_{ij}$ between pairs of loci 
to phylogenetic effect when $i$ or $j$ for one of the three Variants of Concern (VoC) `alpha'~\cite{B.1.1.7}, `beta'~\cite{VoC-6,Tegally2021} or `delta'~\cite{Indian_variant}. 
The corresponding tables of mutations and time evolution of mutation frequencies
were recently reported in~\cite{Zeng2021b}.

\subsection{Fraction of residual couplings over rank}\label{ss-FRC}
Let us define the fraction of residual couplings $R$ over $n$ as follows. We start by ranking all possible couplings by their score $C_{ij}$ or $J_{ij}$ computed as in Eq.(\ref{eqn:Frobenius_norm}).
Within the $n$th highest ranked couplings, a number $k$ of them is removed when it is likely not to be due to an underlying epistatic effect, in particular when\\
1. at least one of the extrema, \textit{i.e.}, $i$ or $j$ in $C_{ij}$ / $J_{ij}$ is located in non-coding regions; \\
2. the extrema are too close ($|i-j|\le5$ bps;); \\
3. the terminals, $i$ or $j$ of $C_{ij}$ or $J_{ij}$, are in focused VoCs.\\

The fraction of residual couplings is then defined as
\begin{equation}
    R(n) = \frac{n-k}{n}.
\label{eqn:fraction_retained_links}
\end{equation}

We compute it to both the inferred couplings by plmDCA and correlation analysis. As shown in Fig. \ref{fig:fraction_of_remain_links}, the curve for residual couplings defined by plmDCA lies well above the curve defined by correlations for small values of rank $n$. 
This hints at the fact that the leading couplings $J_{ij}$ are more likely to capture epistatic effects than leading correlations $C_{ij}$, since an higher number of the latter is removed according to the above criteria for the same $n$. We will comment more on this in Sec.(\ref{ss-LEAD-COR}).

\section{Results\protect}

\subsection{The genome-wide variability (GWV) of SARS-CoV-2 changes in time}

We here define the genome-wide variability (GWV) for a genome as the number of loci that shows variability \emph{i.e.}, the number of loci that survive in the filtering step $L_s$ as described in Sec.(\ref{ss-MSAf}) normalized by the number of genomic sequences $N_{seqs}$.
The threshold $p$ is a free parameter, subject to 
the condition $\left(1-p\right)\cdot N_{seqs}\gg 1$, indicates the expected number of all minor alleles is greater than one.

\begin{figure}[!ht]
\centering
\includegraphics[width=0.9\linewidth]{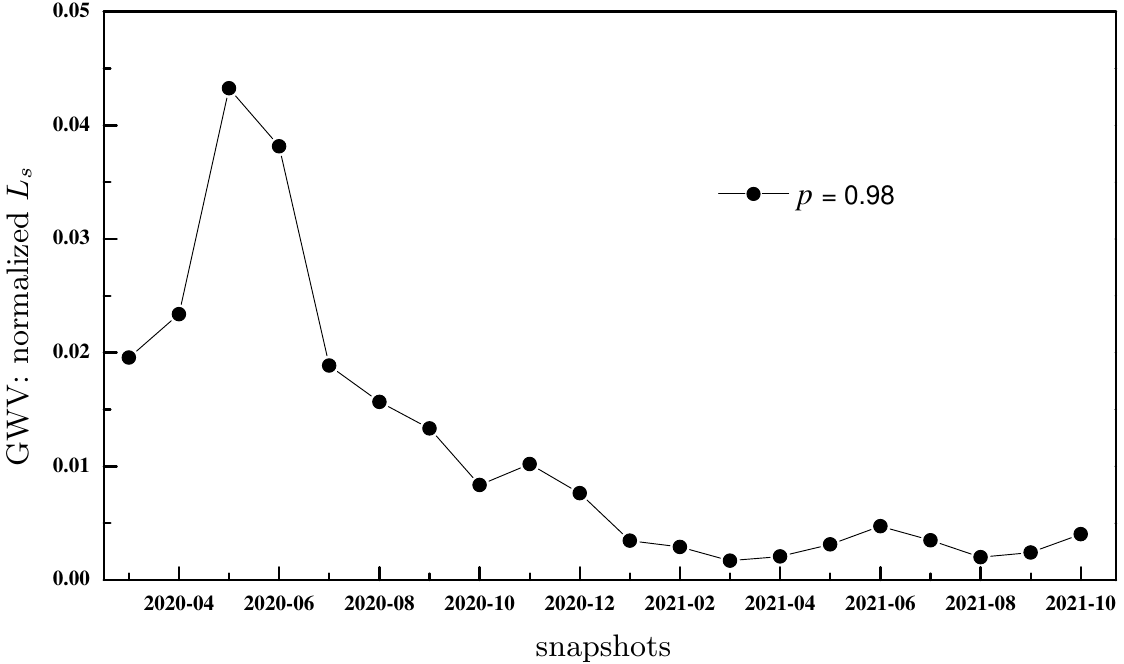}
\caption{GWV, the survived number of loci $L_s$ normalized by $N_{seqs}$ for each MSA with threshold $p=0.98$ until the end of October 2021. At the beginning of the pandemic, GWV increased, however with the increasing number of sequences along month, the GWV roughly decreased.}
\label{fig:num_of_reamin_loci_with_ph_98}
\end{figure}

Fig.~\ref{fig:num_of_reamin_loci_with_ph_98}
shows GWV with threshold $p=0.98$.
The GWV increased in the beginning of the pandemic until 
May of 2020, and then decreased with light fluctuations. 
In the same time period the number of
sequences increased tenfold (Fig.~\ref{fig:seq_num}), thus
the GWV per month has hence decreased.
Similar results hold for 
other choices of $p$ equals
$90\%$ and $99,9\%$ are discussed further in Appendix~\ref{app:num_remain_loci}.

\begin{figure}[!ht]
\centering
\includegraphics[width=0.4\textwidth]{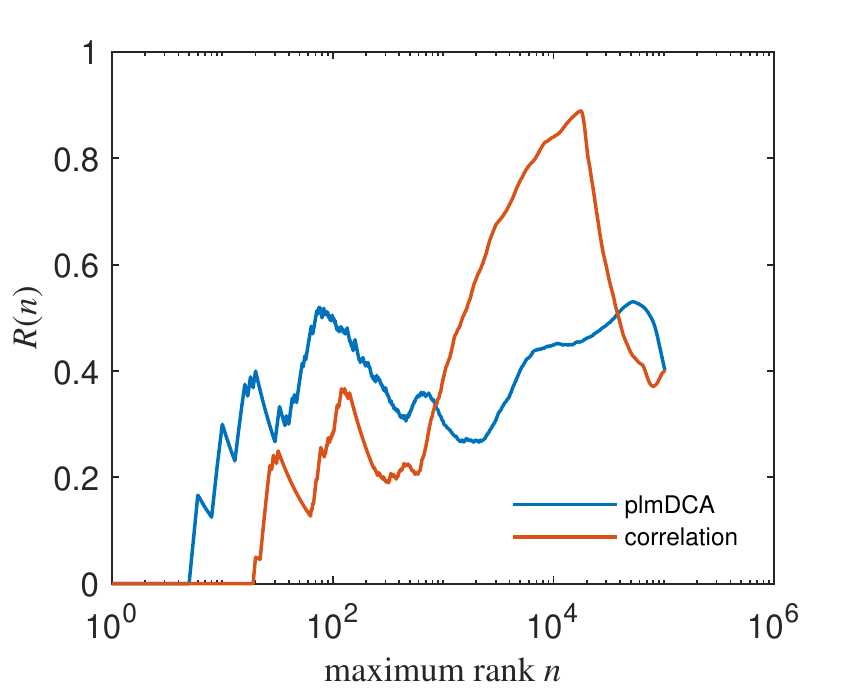}
\put(-37,120){(a)}\\
\includegraphics[width=0.4\textwidth]{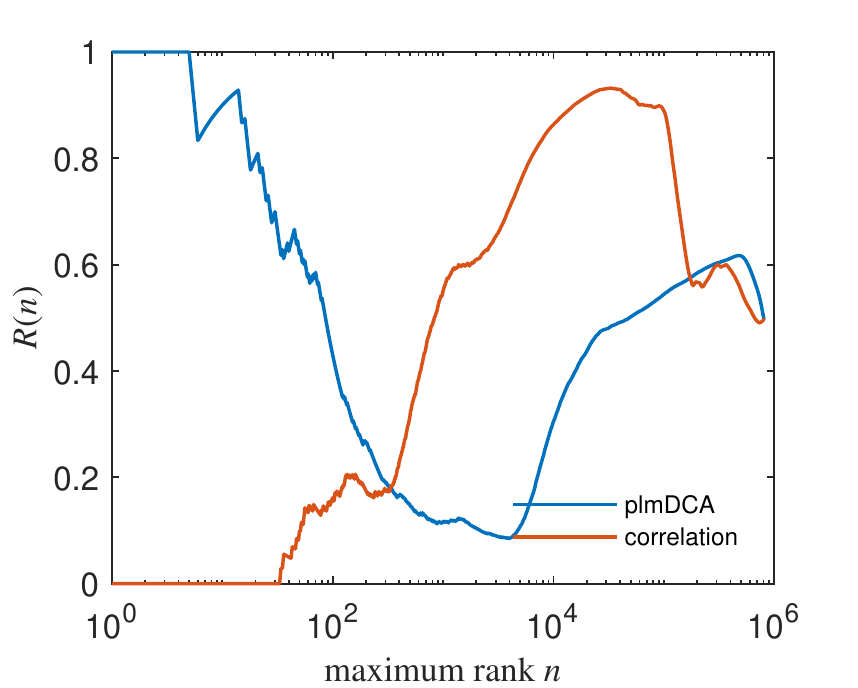}
\put(-37,120){(b)}\\
\caption{Examples of fraction of residual couplings $R(n)$ eq.~\eqref{eqn:fraction_retained_links} as function of maximum rank considered $n$ for October 2020 (a) and 2021 (b). Blue lines show plmDCA $J_{ij}$, red lines show correlations $C_{ij}$. Couplings with $i,j$ satisfying the conditions mentioned in Sec.(\ref{ss-FRC}) have been removed. 
Panel (a) shows the case for the highest 5 ranked plmDCA scores and the highest 20 ranked correlations have been removed. 
Panel (b) shows the case where none of the highest 4 ranked plmDCA scores have been removed while the highest 30 ranked correlations have all been removed. }
\label{fig:fraction_of_remain_links}
\end{figure}
\subsection{Leading correlations can mostly be explained by the growth of focused SARS-CoV-2 VoCs} \label{ss-LEAD-COR}

A simple way to visualize if ranked effects are due to one out of several factors is to plot the contribution of the factor of interest as function of rank. 
A standard procedure in DCA analysis of tables of homologous protein sequences is indeed top-$k$ plots illustrating the fraction of $k$ highest rank predictions which correspond to spatially proximate residue pairs.
For instance, if we want to assess the effect of the rise of variants in the computed couplings, we can simply proceed as described in Sec.(\ref{ss-FRC}) by ranking them by magnitude, taking out those related to the variants and plotting $R(n)$ as in eq.~\eqref{eqn:fraction_retained_links}. 
This is done in Fig.~\ref{fig:fraction_of_remain_links}
for one representative and one exceptional month
(in fact, the only exceptional month,
see Fig.~\ref{fig:Fraction_of_remain_ranks_all_months}
in Appendix~\ref{app:fraction_retained_ranks}. 
In both plots the fractions of highest ranked correlations $C_{ij}$ and 
plmDCA terms $J_{ij}$ are presented,
with part of $C_{ij}$ or $J_{ij}$ being removed when $i$ or $j$ matches the removal conditions listed in (Sec.\ref{ss-FRC}).
The representative month (October 2021,
Fig.~\ref{fig:fraction_of_remain_links}(b))
shows that the leading correlations can mostly
be explained by variations in VoCs alpha, beta and delta.
For the exceptional month (October 2020, Fig.~\ref{fig:fraction_of_remain_links}(a))
the separation between correlations and DCA terms is not clear. 

The essence of the argument is that for the same $n$, leading DCA terms contain much fewer 
pairs where one or both terminals 
appear in the VoCs.
Lists of leading DCA terms are hence, compared to correlations, 
enriched for epistatic interactions. Analogous (and similar) results are shown for the other months in
Appendix~\ref{app:fraction_retained_ranks}.

\begin{figure}[!ht]
\centering
\includegraphics[width=0.49\textwidth]{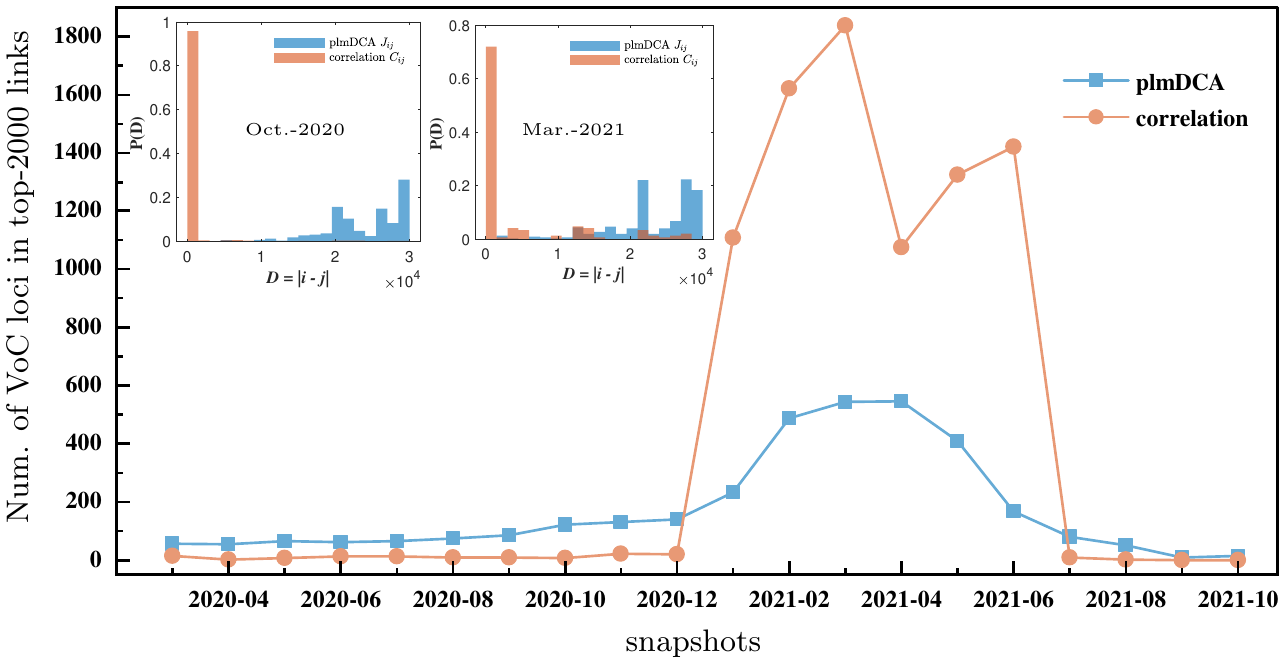}
\caption{Main panel: number of loci located in the focused VoCs from top 2000 $J_{ij}$ (blue squares) and $C_{ij}$ (orange dots) respectively over each month. Correlations provide much more VoC loci than DCAs during the period of December 2020 and July 2021, when the focused VoCs domain. Inner panel: Probability of distance $D=|i-j|$ in top 2000 $C_{ij}$ and $J_{ij}$ for October 2020 and March 2021 respectively. Correlations tend to figure out links with short $D$ both in and out of the VoC domain period. In the main panel, the links with $i$, $j$ satisfying the 1st and 2ed removal conditions are discarded. For the inner panel, the links meet all removal conditions are not considered.}
\label{fig:Num_loci_in_VOCs}
\end{figure}

To check the other possible confounding effects, the number of loci that appears in the focused VoCs are counted in the top 2000 $C_{ij}$ and $J_{ij}$ over month, as presented in the main panel of Fig. \ref{fig:Num_loci_in_VOCs}. It is clearly shown that correlations containing much more VoC loci comparing with that from DCAs during the pandemic period of the focused VoCs. 
Meanwhile, the distributions of distance $D$ between $i$ and $j$ in these tops are provided in the inner panels of Fig. \ref{fig:Num_loci_in_VOCs} for October 2020 and March 2021 respectively.
In both cases, correlations tend to figure out links with short $D$. This explains the big jump between $C_{ij}$ and $J_{ij}$ during the end of 2020 to the middle of 2021.

\subsection{Inferred epistasis has both invariant and variant aspects} 
\begin{figure}[htpb]
\centering
\includegraphics[width=0.47\textwidth]{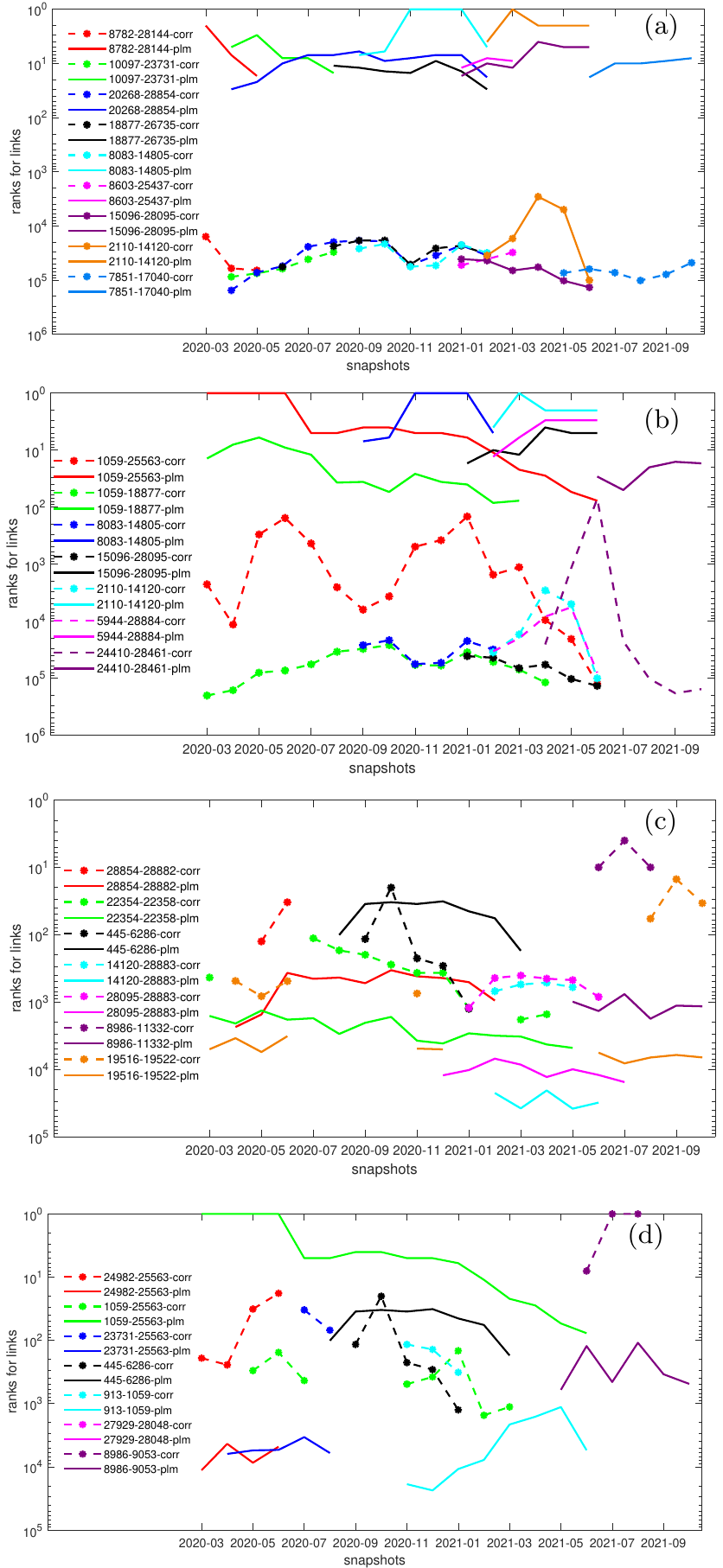}
\caption{Ranks for a subset of residual epistatic interactions per month inferred by plmDCA (solid lines) $J_{ij}$ and correlations  (dashed lines) $C_{ij}$. 
The pairs with $i$ or $j$ matching condition 1 and 2 described in Sec. (\ref{ss-FRC}) have been discarded.
(a) and (b) display the ranks of a subset of top-200 $J_{ij}$ and their counterparts $C_{ij}$ with $i$, $j$ excluded and included in the focused VoCs respectively.$J_{ij}$ extend more or less the same period with their counterparts but with much higher rank.
Similarly, the ranks of a subset of top-2000 $C_{ij}$ and the corresponding $J_{ij}$ with $i$ or $j$ located out of or inside the focused VoCs are presented in (c) and (d) respectively. $J_{ij}$ extend much longer than $C_{ij}$ over time.The data for this figure is given in Datasets 2 and 3 of Appendix~\ref{app:datasets}.}
\label{fig:ranks_plm_200_plmDCAs_vs_correlation}
\end{figure}

One novelty of the analysis presented in this work
is that the dataset is much larger than in previous 
contributions~\cite{Zeng2020pnas,CRESSWELLCLAY2021},
and that it has been grouped monthly by sampling time.
A second novelty is that phylogenetic confounders 
have been eliminated by excluding inferred links 
where one or both loci appear in 
 the focused VoCs of SARS-CoV-2. 
Fig.~\ref{fig:ranks_plm_200_plmDCAs_vs_correlation}
displays the ranks of leading residual epistatic 
interactions $J_{ij}$ (solid lines) and correlations $C_{ij}$ (dashed lines) with same $i$ and $j$ as a function of sampling time.  
A subset of top 200 $J_{ij}$ with $i$s or $j$s excluded or included in the focused VoCs are provided in  Fig.~\ref{fig:ranks_plm_200_plmDCAs_vs_correlation}(a) and (b) respectively. Their counterpart $C_{ij}$ are shown in dashed lines. One feature that stands out on these two sub-graph is that as long as they appear in the data, both types of ranks appear fairly stable, but the ranks of correlations is far lower. Similarly, a subset of top 2000 $C_{ij}$ and their corresponding $J_{ij}$ with $i$ or $j$ located out of or inside the focused VoCs are displayed in Fig.~\ref{fig:ranks_plm_200_plmDCAs_vs_correlation}(c) and (d) respectively. Here the $J_{ij}$s last longer than the corresponding $C_{ij}$s over time.

Furthermore Fig.~\ref{fig:ranks_plm_200_plmDCAs_vs_correlation} shows that none of the interactions appear for the entire period, but only in some time window.
Outside this window, the frequency of the major allele
of one or both loci in a pair rises above the threshold $p$ and is hence discarded because of lack of variability. As a consequence, the pair hence disappears from the analysis.
We can therefore at best have a \textit{transient} QLE phase, as defined in Sec.(\ref{subsec:transient_QLE}).

\subsection{A subset of mutations of Variant of Concern omicron has non-trivial dynamics} 
\begin{figure*}
\centering
    \includegraphics[width=0.7\textwidth]{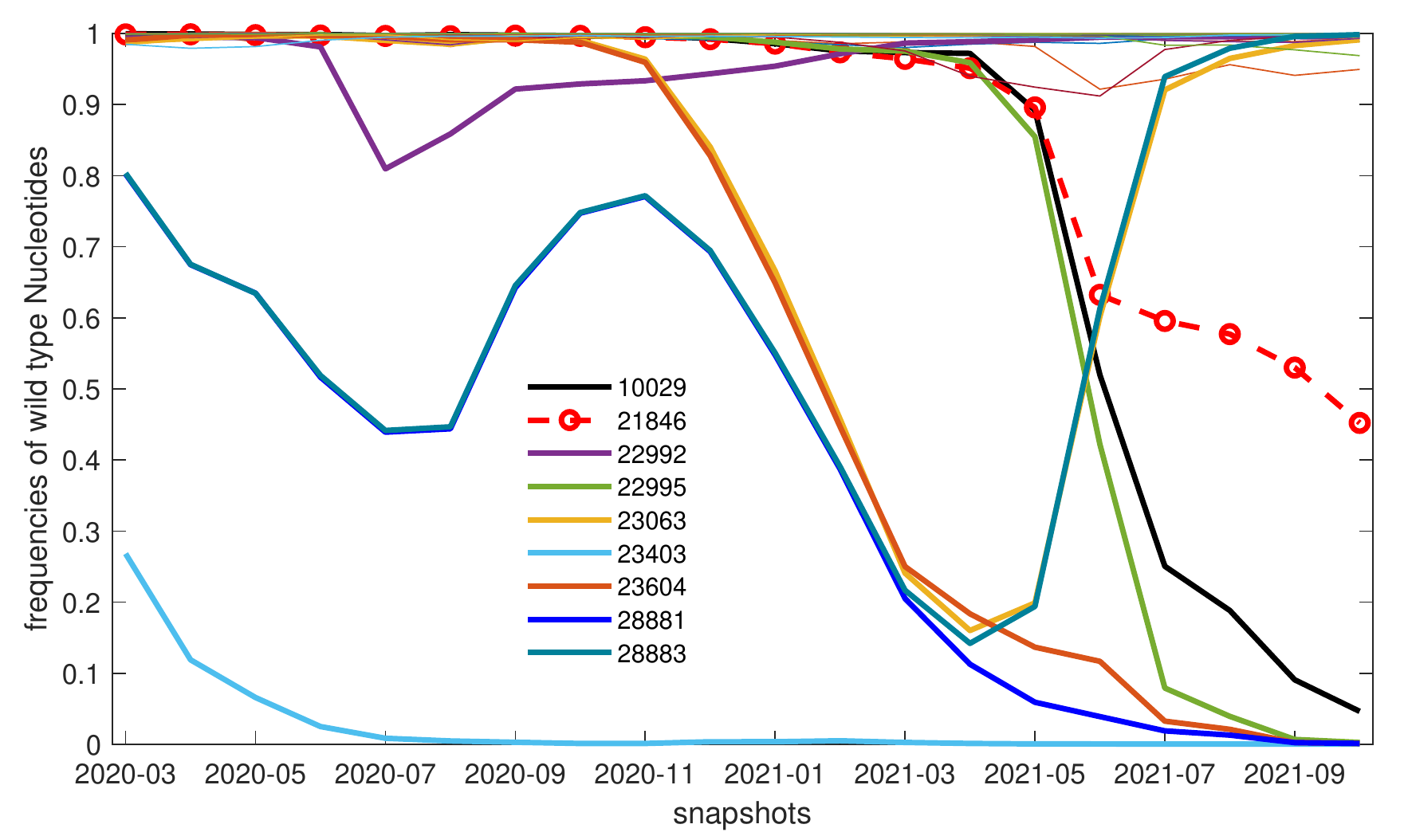}
    \caption{Wild-type nucleotide frequencies 
    of the loci of the omicron variant listed in~\cite{Omicron}. The wild type is defined here as reference sequence ``Wuhan-Hu-1", GISAID access ID: EPI\_ISL\_402125; nucleotides are numbered based on their locus in the latter. 
    The growth of the omicron variant happens later than the time-window displayed.
    For most of the omicron loci the corresponding wild-type frequency remains 
    close to $1$ (thin lines in figure). 
    Loci showing large fluctuations are plotted with bold lines. 
    $23403$ is the Spike mutation  
    D614G which rose early in
    the pandemic while
    $23063$ and $23604$
    are S:N501Y and S:P681H,
    which appeared in alpha.
    $22995$ is S:T478K  which appeared in delta.
    The remaining mutations are $21846$ (dashed marked red line) discussed in text, $28881$ (N:R203M) which is listed in delta but follows another dynamic~\cite{Zeng2021b}, $10029$ which follows a trajectory characteristic of delta without being listed for it, and $28883$ which does the same for alpha.} 
    \label{fig:Nuc_fre_Omicron}
\end{figure*}

Somewhat out of line with the main thrust of
the presentation, we also find it of interest to 
describe the dynamics of the individual mutations 
listen for VoC omicron. To our knowledge these
observations have not been made previously in the literature.
In earlier work we showed that a subset of the mutations 
in alpha, beta and delta have different dynamics
than would be expected from a clone growing de novo~\cite{Zeng2021b}.
In Fig.~\ref{fig:Nuc_fre_Omicron}
we show that the same holds for the more-recently characterized
variant omicron~\cite{Omicron}.
The red-dot trajectory shows the
frequency of a nucleotide substitution
at position 21846, which corresponds to S:T95I in Spike, the latter being listed as one of the defining mutations for the omicron variant. The string S:T95I indicates the amino acid substitutions: in this case it means that the point mutation occurred at the locus 21846 causes in the translated polypeptide chain of the Spike (S) a mutation of the amino acid 95 from T (threonine) to I (isoleucine).
However, this mutation in the
\texttt{N}-terminal domain of S1 subunit of the Spike protein
rose quickly in frequency from May-21 to June-21
in GISAID database, and has since been found
in about half of the samples.
Its prevalence therefore cannot be explained
by the rise of omicron, which appeared later.
Indeed, T95I 
was among the most 
common in the variant B.1.526 (Iota)
which spread mainly in USA in late 2020
and early 2021~\cite{West2021}, and
has been observed in strains
classified as VoC delta
circulating in France~
\cite{Verdurme2021},
and in UK and
Germany~\cite{Rono2021}.
It is therefore an example of a mutation 
which though classified as a defining mutation
of one strain of the virus, in fact is more 
widely spread, and can be found also in other strains.
The other curves in Fig.~\ref{fig:Nuc_fre_Omicron}
with a dynamics different from omicron can
mostly be explained as also belonging to 
VoCs alpha and delta.

\subsection{DCA detects epistatic interactions between loci in Spike, and between Spike and other genes} 
Let us focus here on the couplings involving sites on the SARS-CoV-2 that code for the Spike protein, crucial for the virus to bind to target cells. In an earlier study using data up to August 2020,
only one large plmDCA score involving a locus in
Spike was detected, at genomic position 23403
\cite{Zeng2020pnas} (Table 1).
This \texttt{G}$\to$\texttt{A} substitution was deemed to be due to a phylogenetic effect as detected 
by a randomization procedure
\cite{Zeng2020pnas} (Table 2), and therefore not
retained as a predicted epistatic link in that study. 
In fact, this substitution 
corresponds to the well-known mutation Spike D614G
which grew in frequency in the early
phase of the pandemic.

In the present (larger) month-by-month data we find several persistent plmDCA couplings with terminals in the Spike coding regions. Some of them are related to the variants alpha, beta, delta and also the more recent omicron~\cite{Omicron} (in the sense above), some other are not. Table~\ref{table:things-in-Spike-both} gives for the months August-October 2021 the couplings whose both terminals are in a Spike coding region while Table~\ref{table:things-in-Spike} lists those for which only one of the extrema is in a Spike coding region; corresponding data for all months is mentioned in Appendix~\ref{app:datasets} Dataset 4.

The largest inferred Spike-Spike interaction in all
three months, not related to any mutation appearing in alpha, beta, delta or omicron,
is between two synonymous mutations $D574D$ and $D1259D$.
Most of the other pairs in Table~\ref{table:things-in-Spike-both}
either have one terminal listed in delta, or are somewhat close along the genome,
within $35$ bp, or involve a synonymous
mutation.
This includes
the pair $(21846,24208)$ appearing in October, where 
the first locus 
is the mutation S:T95I, part of the definition of omicron and
discussed above, while
the second locus is S:I882I.

The first prediction appearing in Table~\ref{table:things-in-Spike},
in all three months, is the pair $(17236,24208)$
where the first locus is in gene nsp13
and the second is a synonymous mutation
in Spike. This is fact the largest
effect detected by the DCA analysis
in all three months (rank 1 in
Table~\ref{table:things-in-Spike}).

The second prediction in
Table~\ref{table:things-in-Spike},
ranked respectively
14, 13 and 10,
is the pair $(7851,21846)$
where the first is the mutation A1711V
in nsp3.
In Table~\ref{table:things-in-Spike}
the variation at locus $21846$ (S:T95I)
appears also together with
nsp2:K81N in all three months,
together with 
ORF8:P38P and
ORF7a:V71I  in August,
and together with 
nsp6:A2V, 
N:S327L, 
nsp13:I334V, 
nsp12:S837S, 
ORF3a:E239, 
N:Q9L
and
ORF7a:G38G 
in October.
The mutation nsp3:1711V
was a defining mutation
of Variant of Interest labelled N.9
discovered in Brazil in 2020
\cite{Resende2021}.
The mutation nsp2:K81N
has been detected in variants of
VoC
delta circulating in Russia
\cite{Klink2021}.

The third prediction in
Table~\ref{table:things-in-Spike},
ranked respectively
20, 16 and 17,
is the pair $(28461,24410)$
where the first is the mutation G63D
in N, and the second is N950D in Spike.
N:G63D is a defining mutation
of VoC delta while
S:N950D is a reverse mutation of
delta defining mutation
D950N, identified
as such in a recent study~\cite{Rono2021}.

The next two predictions which appear in all three
months and which do not involve any of the above or any variants
both involve
locus 21897 (S:S112L)
with partners respectively
26107 (ORF3a:E239Q)
at ranks 52, 57 and 60,
and 
27507 (ORF7a:G38G)
at ranks
57, 71 and 74.
Spike mutation S112L
was recently associated to vaccination breakthrough infections in New York City~\cite{Duerr2021}.
That study also identified that genes ORF3a (56\%) and ORF8 (67\%) had higher numbers of sites with enriched mutations in breakthrough sequences. 
ORF3a mutation E239Q is located at the protein C-terminal,
and appears in the sub-variant of
VoC delta variously labelled
 AY.25 and B.1.617.2.25;
it has no annotation in UniProt.

\begin{table*}[htpb]
\small
\setlength\tabcolsep{0.5pt}
\begin{ruledtabular}
\begin{tabular}{lllll|lllll|lllll}
\multicolumn{5}{c}{\textbf{\large August 2021}} & \multicolumn{5}{c}{\textbf{\large September 2021}} & \multicolumn{5}{c}{\textbf{\large October 2021}} \smallskip\\
rank & locus 1 & AA-m. & locus 2 & AA-m. & rank & locus 1 & AA-m.& locus 2 & AA-m. & rank & locus 1 & AA-m.&  locus 2  & AA-m.  \\
\hline
7  & 23284 & D574D & 25339 & D1259D & 7 & 23284 & D574D & 25339 & D1259D & 9  & 23284 & D574D & 25339 & D1259D \\
16 & 21987 & G142D & \cellcolor{green!70}24410 & D950N  & 15& 21987 & G142D & \cellcolor{green!70}24410 & D950N  & 11 & 21995 & T145H & 22227 & A222V  \\
67 & 22093 & M177I & 22104 & G181V  & 45& 21995 & T145H & 22227 & A222V  & 15 & 21987 & G142D & \cellcolor{green!70}24410 & D950N   \\
70 & \cellcolor{green!70}22917 & R452L & \cellcolor{green!70}22995 & K478T  &   &       &	&       & 	 & 135& \cellcolor{red!70}21846 & T95I  & 24208 & I882I   \\
71 & 22082 & P174S & 22093 & M177I  & 	&	&	&	&        &    &       &       &           &  \\
74 & 22081 & Q173H & 22093 & M177I  &   &	& 	&	& 	 &    &	      &       &	          &  \\
190& 22082 & P174S & 22104 & G181V  &	&       &       &       & 	 &    &	      &       &	          &  \\
195& 22081 & Q173H & 22104 & G181V  &   &       &	&	&        &    &	      &       &	          &   \\
\end{tabular}
\caption{\footnotesize Largest DCA terms with both terminals in Spike coding region, August-October 2021. Top-200 couplings computed as plmDCA scores are considered. For each of them in the three months displayed, there's the indication of the rank, the two loci involved and the corresponding amino acid (AA) mutations. 
Green color indicates that this mutation is found in delta variant. Red color indicates that this mutation is found in omicron variant. Couplings with one or both terminals colored green are attributed to a phylogenetic effect. The single pair with one terminal colored red is not attributed to a phylogenetic effect, the growth of omicron being later than October-2021. Omicron mutations used here are taken from~\cite{Omicron} on page 18, deletions not considered.}
\end{ruledtabular}
\label{table:things-in-Spike-both}
\end{table*}

\begin{table*}
\small
\setlength\tabcolsep{0.5pt}
\begin{ruledtabular}
\begin{tabular}{llllr|llllr|llllr}
\multicolumn{5}{c}{\textbf{\large August 2021}} & \multicolumn{5}{c}{\textbf{\large September 2021}} & 
\multicolumn{5}{c}{\textbf{\large October 2021}} \smallskip\\
rank & \multicolumn{2}{c}{Partner} & \multicolumn{2}{c}{Spike} & rank & \multicolumn{2}{c}{Partner} & \multicolumn{2}{c}{Spike} & rank & \multicolumn{2}{c}{Partner} & \multicolumn{2}{c}{Spike} \\
                      & locus          & \multicolumn{1}{c}{AA-m.}        & locus         & \multicolumn{1}{c}{AA-m.}        &                       & locus          & \multicolumn{1}{c}{AA-m.}        & locus         & \multicolumn{1}{c}{AA-m.}       &                       & locus          & \multicolumn{1}{c}{AA-m.}        & locus         & \multicolumn{1}{c}{AA-m.}    \\

\hline
1  & 17236 & nsp13:I334V & 24208 & I882I     & 1  & 17236   & nsp13:I334V & 24208 & I882I  &  1  & 17236  & nsp13:I334V   & 24208 & I882I	 \\ 
14 &  7851 & nsp3:A1711V &  \cellcolor{red!70}21846 & T95I      & 13 &  7851   & nsp3:A1711V &  \cellcolor{red!70}21846 & T95I   &  10 &  7851   & nsp3:A1711V  &  \cellcolor{red!70}21846 & T95I	  \\
20 & 28461 & N:G63D      &  \cellcolor{green!70}24410 & D950N     & 16 & 28461   & N: D63G     &  \cellcolor{green!70}24410 & D950N  &  17 & 28461  & N:D63G	  &  \cellcolor{green!70}24410 & D950N     \\               
27 &  1048 & nsp2:K81N   &  \cellcolor{red!70}21846 & T95I      & 36 &  1048   & nsp2:K81N   &  \cellcolor{red!70}21846 & T95I   &  20 & 25614  & ORF3a:S74S    & 21995 & T145H     \\	      
52 & 26107 & ORF3a:E239Q & 21897 & S112L     & 52 & 25614   & ORF3a:S74S  & 21995 & T145H  &  21 & 25614  & ORF3a: S74S   & 22227 & A222V     \\	      
57 & 27507 & ORF7a:G38G  & 21897 & S112L     & 57 & 26107   & ORF3a:E239Q & 21897 & S112L  &  30 &  1048   & nsp2:K81N	  &  \cellcolor{red!70}21846 & T95I	  \\       
62 & 18086 & nsp14:T16I  & 22792 & I410I     & 58 & 25614   & ORF3a:S74S  & 22227 & A222V  &  51 & 10977  & nsp6:A2V      &  \cellcolor{red!70}21846 & T95I	   \\       
76 & 27291 & ORF6:D30D   & 24208 & I882I     & 71 & 27507   & ORF7a:G38G  & 21897 & S112L  &  56 & 27291  & ORF6:D30D	  & 24208 & I882I	   \\  
79 &  1729 & nsp2:V308V  & 22792 & I410I     & 82 & 27291   & ORF6:G30G   & 24208 & I882I  &  60 & 26107  & ORF3a:E239Q	  & 21897 & S112L	  \\              
151& 28007 & ORF8:P38P   &  \cellcolor{red!70}21846 & T95I      & 83 & 11514   & nsp6:T181I  & 22227 & A222V  &  63 & 29253  & N:S327L       &  \cellcolor{red!70}21846 & T95I	   \\         
168& 27604 & ORF7a:V71I  &  \cellcolor{red!70}21846 & T95I      & 128& 17236   & nsp13:I334V &  \cellcolor{red!70}21846 & T95I   &  64 & 18744  & nsp14:T235T	  &  \cellcolor{red!70}24130 & N856N \\                 
174& 17236 & nsp13:I334V &  \cellcolor{red!70}21846 & T95I      & 151& 18744   & nsp14:T235T &  \cellcolor{red!70}24130 & N856N  &  74 & 27507  & ORF7a:G38G    & 21897 & S112L \\	      
197& 11514 & nsp6:T181I  & 22227 & A222V     & 190& 5584    & nsp3:T955T  & 22227 & A222V  &  80 & 17236  & nsp13:I334V	  &  \cellcolor{red!70}21846 & T95I  \\                
   &    &          &    &        & 195& 13019   & nsp9:L112L  & 22227 & A222V  &  124& 15952  & nsp12:S837S	  &  \cellcolor{red!70}21846 & T95I  \\                
   &    &          &    &        & &        &         &    &   &  153& 26107  & ORF3a:E239    &  \cellcolor{red!70}21846 & T95I\\                      
   &    &          &     &         &  &       &         &    &   &  163& 28299  & N:Q9L         &  \cellcolor{red!70}21846 & T95I	  \\          
   &     &           &     &        &   &         &         &   &  &  190& 27507  & ORF7a:G38G    &  \cellcolor{red!70}21846 & T95I \\	              
   &     &           &     &        &  &         &         &   &   &  194& 11562  & nsp6:C197F	  & 21897 & S112L \\	              
   &     &           &     &        &   &        &          &   &   &  197& 11514  & nsp6:T181I	  & 22227 & A222V \\
\end{tabular}
\caption{\footnotesize Largest DCA terms with only one terminal in Spike coding region, August-October 2021. 
Top-200 couplings computed as plmDCA
scores are considered. For each of them in the three months displayed, there’s the indication of the rank, the locus in the Spike coding region and corresponding amino acid (AA) mutation, the locus in the partner coding region and corresponding amino acid (AA) mutation. 
Green color indicates that this mutation is found in delta
variant. Red color indicates that this mutation is found in omicron
variant.
Pairs with one or both terminals colored green are attributed 
to a phylogenetic effect, while the several
pairs with one terminal colored red 
are not, the growth of omicron
being later than October-2021.
Omicron mutations used here are taken from~\cite{Omicron} on page 18, deletions not considered.
}
\end{ruledtabular}
\label{table:things-in-Spike}
\end{table*}

\section{Discussion}
In this work we have applied the Direct Coupling Analysis (DCA)
methodology~\cite{Weigt2009pnas,MorcosE1293,Hopf-2012a,Ekeberg-2013a,Ekeberg-2014a,Nguyen2017} 
to identify putative epistatic interactions between 
pairs of loci in the SARS-CoV-2 virus.
We have described the rationale for such an approach
based on the Quasi-Linkage Equilibrium (QLE) mechanism of
Kimura~\cite{Kimura1965,NeherShraiman2011},
which we have recently combined with DCA in 
\textit{in silico} validation~\cite{Gao2019,Zeng2020,Zeng-Mauri-2021}.
As part of the world-wide effort to combat the COVID19
epidemic an unprecedented number of genomes of the 
disease agent have been obtained and released 
through open repositories. In this study we have thus
been able to use more than
three and a half million full-length high-quality 
SARS-CoV-2 genomes from GISAID deposited until October 2021~\cite{GISAID}.
Such very large, quasi-exact and easily accessible
data
resources will very likely be the norm in future pandemics.
Methods to turn them into actionable information in new ways 
are therefore of high relevance.
Except for the more that order of magnitude larger 
data size, the main methodological novelty in this study 
has been to separate genomes as to sampling date (by month).
We have hence been able to carry out a \textit{temporal} epistasis inference, to the best of our knowledge for the first time.

Our main finding is that the leading terms identified
by DCA are more stable over time than correlations. This
is an argument in favor of the global SARS-CoV population
exhibiting characteristics of QLE, as would be expected 
from the substantial rate of recombination characteristic
of coronaviruses~\cite{LaiCavanagh1997} and the sometimes
high rate of circulating infections in the human population world-wide.
This finding however comes with a caveat: DCA analysis (and correlation analysis) is necessarily based on observed variability which disappears if an allele at a locus is lost. This is indeed also what we find. The stability of DCA terms therefore only pertain for the time window when the mutations at both terminals appear in a significant proportion of the samples. Few of the epistatic interactions found in two earlier studies~\cite{Zeng2020pnas,CRESSWELLCLAY2021} are in fact found in the later data, as one or both of the corresponding mutations have either since been lost or reached fixation. 

We refer to the resulting setting \textit{temporal epistasis inference}.
In earlier theoretical work we identified the possibility
of retrieving epistatic parameters from pairwise variations in
a population even though single-locus frequencies vary greatly
\cite{Gao-github}.
In this work we have found that such an effect appears in
data, and is reflected in the epistasis prediction
pipeline through the appearance and/or disappearance of
predicted pairs.
The biological relevance is that epistasis can be detected in 
a transient phase, and then used as input to further analysis at a later time,
when variations at one or both terminals will have disappeared, and epistasis can
no longer be detected from the sequences present in the population.
We further remark that in the data at hand
(SARS-CoV-2 sequences collected in the COVID-19 pandemic)
evolutionary parameters are themselves most likely 
changing with time. The most immediate effect is the 
changed fitness landscape (to the virus) after large-scale
vaccination (of the human hosts). We have in this work not tried
to estimate such effects.

The main success story of DCA applied to biological data has been to predict spatial residue-residue contacts in proteins~\cite{Cocco2018}.
In that important application 
accuracy of predictions can be assessed by comparing to distance data in resolved protein structures. 
Spatial proximity is the main mechanism behind and a 
relevant proxy for epistasis within one gene (one protein).
It is a general feature of DCA that the 
accuracy is generally highest for the largest predictions,
typically visualized through plots of the True Prediction Rate
of $k$'th largest predictions ($TPR(k)$)\cite{Cocco2018}.
On the global genome scale labelled test data of the same kind
is not available, and evaluation will necessarily 
be in terms of potential biological or medical relevance,
compared to literature, or other data.

In the bacterial domain,
in an earlier study based on around 3,000 full-length genomes
of the bacterial pathogen 
\textit{Streptococcus pneumoniae} we hence found as main terms epistatic interactions 
between loci in the PBP family of proteins
central to antibiotic resistance in the pneumococcus \cite{Skwark-2017a};
analogous results have also been found 
for the gonococcus~\cite{Schubert2019}.
Recent results 
using on the hand than $60,000$  
\textit{Escherichia coli} genomes, 
and on the other a set of closely related 
other bacterial genomes,
lead to testable predictions on amino acid variability
\cite{Vigue2022}.

In the viral domain
DCA methods have been applied to the genes
coding for the envelope
in a well-known series of papers
\cite{FERGUSON2013,Chakraborty2013pre,Chakraborty2016pre,Chakraborty2017Ropp,Chakraborty2018pnas,barton2019modelling};
more have led to experimental tests
\cite{Chakraborty2021pnas} promising for anti-viral drug
and vaccine development.
The same group has also extended the analysis 
to polio \cite{Chakraborty2020nc}.
On the global genome level a recent contribution used
whole-genome sequences of a set of coronaviruses to 
predict mutability using DCA methods which were then
assessed by the use of the same GISAID data base as
we have used here
\cite{Rodriguez-Rivas2022}.
Leveraging a more variable set of genomes 
is an alternative and possibly
more robust avenue to obtain biologically viable
predictions than the route taken here; the issue 
however merits further investigations.

We have here limited ourselves
to a discussion of the top-200 predictions per month that are
also stable in rank over the last three months of data
(August-October 2021), and which involve loci in Spike.
We find several DCA terms associated
to Variants of Concern delta and omicron, which we in
the case of delta attribute to a phylogenetic effect.
On methods to remove phylogeny as a confounder of DCA
we refer to~\cite{RodriguezWeigt-2021,Horta2021},
and as described in our earlier contribution
\cite{Zeng2020pnas}.
The most prominent of the mutations in omicron is S:T95I at genomic position 21846.
Although a defining mutation for this VoC, it was actually found in approximately
half of the genomes collected world-wide in the time period 
August-October 2021. The inferred epistatic interactions between 
S:T95I and loci in other genes are hence examples 
of interactions that were detectable in data up to
the end of 2021, but which is not be detectable anymore 
as the omicron variant has taken over fully.

Our results of potential biological 
and medical relevance are 
given in
Table \ref{table:things-in-Spike-both}
for epistatic interactions between two loci
out of which at least one in Spike.
We surmise that the most interesting of those
are two epistatic interactions involving
Spike mutation S112L, recently shown to be
associated to vaccination
breakthrough infections~\cite{Duerr2021}.
One of its interaction partners is
mutation ORF3a:E239Q where
ORF3a is a cation channel
protein unique to the 
coronavirus family~\cite{Kern-2020}
and known to be involved in
inflammation of lung tissue 
and severe
disease outcomes~\cite{Lu2006,Siu2019,Ren2020}.
In the earlier study~\cite{Zeng2020pnas}
several other mutations in ORF3a 
appeared prominently; in this study
a new one does so together with 
a mutation in Spike.

\begin{acknowledgments}
We thank Dr Edwin Rodr{\'i}guez Horta, Profs Martin Weigt and Roberto Mulet for numerous discussions. 
EA thanks Kaisa Thorell and Rickard NordR{\'e}n for useful suggestions.
The work of HLZ  was sponsored by NSFC 11705097, NY221101. 
YL was supported by KYCX21-0696.
The work of EA was supported by the
Swedish Research Council grant 2020-04980.
\end{acknowledgments}

\appendix

\section{The number of sequences sampled per month has increased during the pandemic} 
\label{app:num_seq_per_month}
\begin{figure}[!ht]
\centering
\includegraphics[width=0.9\linewidth]{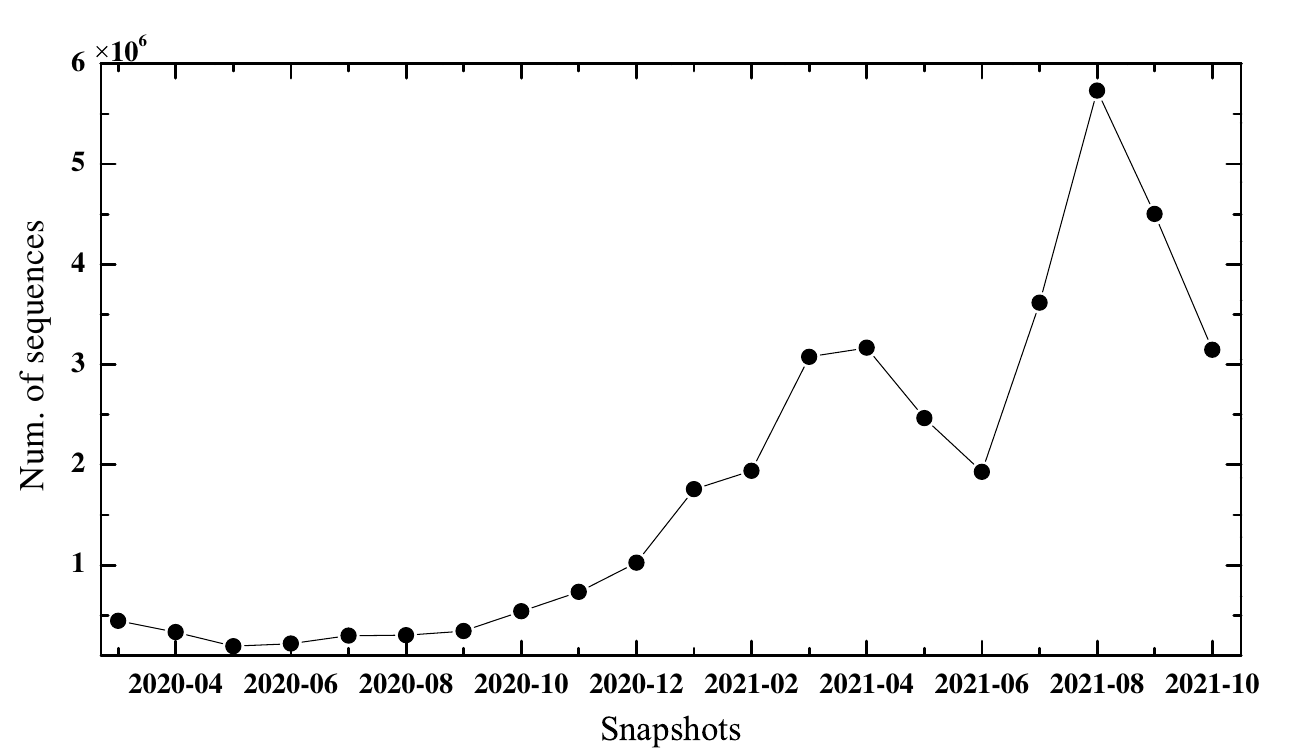}
\caption{Number of complete and high quality SARS-CoV-2 sequences deposited in the GISAID repository per month until the end of October 2021, stratified by
month of sampling time. The accession IDs of the samples from GISAID used in this work are given in
Appendix~\ref{app:datasets} Datasets 1.}
\label{fig:seq_num}
\end{figure}

Fig.~\ref{fig:seq_num} shows the number of whole-genome 
high-quality SARS-CoV-2 sequences deposited in GISAID 
and stratified by month. 
With some irregularity this number has grown exponentially 
since the summer of 2020, and is now around
half a million SARS-CoV-2 genomes per month.

\section{GWV with other filtering thresholds}
\label{app:num_remain_loci}
\begin{figure}[!ht]
\centering
\includegraphics[width=0.85\linewidth]{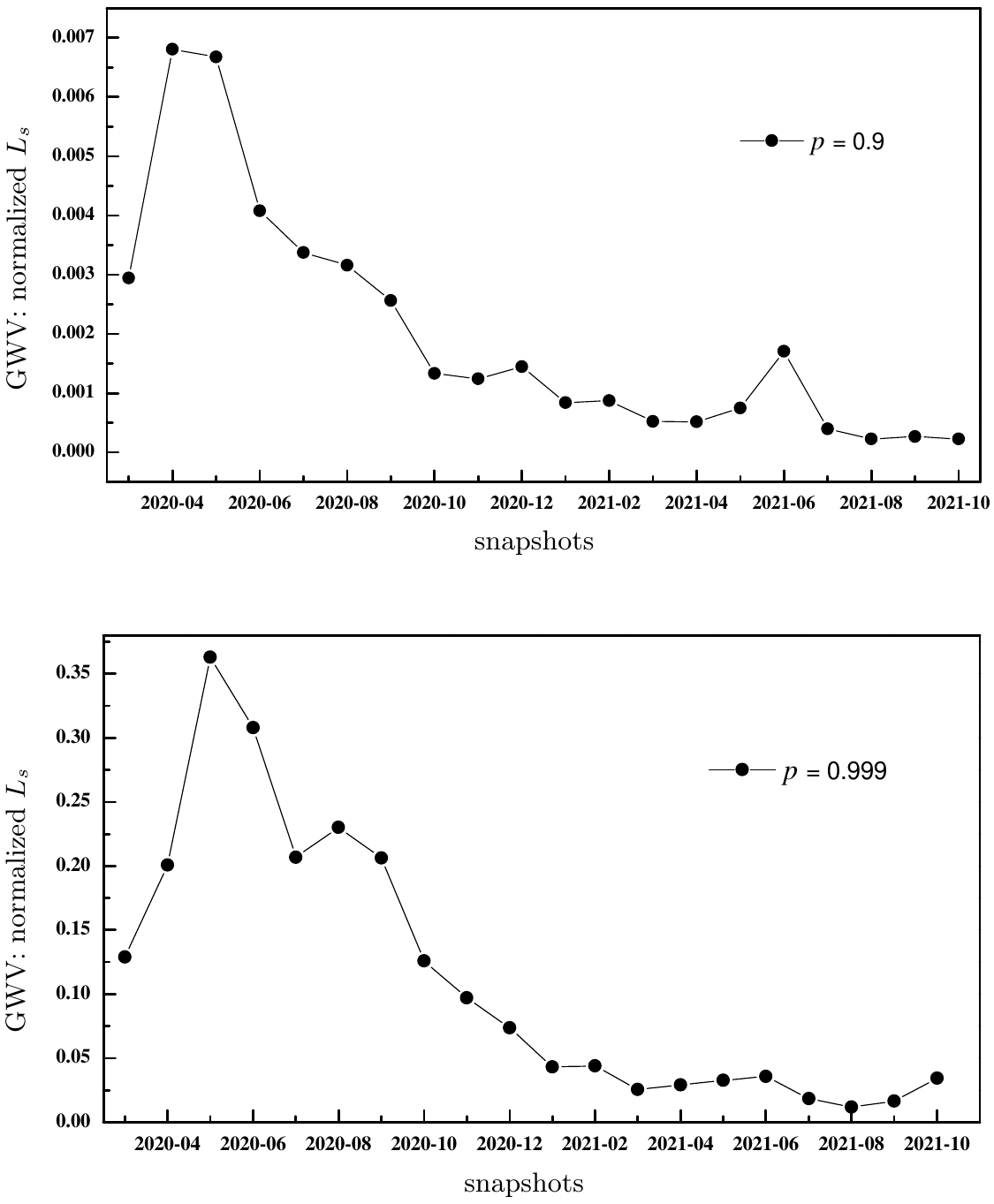}
\caption{Normalized survived loci $L_s$ by the total number of sequences $N_{seqs}$ per month with $p=0.9$ (upper) and $p=0.999$ (bottom). In our analysis, loci (MSA columns) where any of the nucleotides is found with a frequency greater than $p$ is excluded. }
\label{fig:num_of_reamin_loci_with_ph_90}
\end{figure}

We computed the frequencies of nucleotides along each locus/column in each MSA matrix. If the frequency of any of the nucleotides is larger than the given value of $p$, this locus will be excluded in the following epistasis analysis. To complement Fig.~\ref{fig:num_of_reamin_loci_with_ph_98} in the main text, we show plots for the normalized number of survived loci $L_s$ by the number of sequences $N_{seqs}$ in each MSA with different values of $p$ here in Fig.  \ref{fig:num_of_reamin_loci_with_ph_90}. The upper panel is for $p=0.9$ while the bottom one for $p=0.999$ respectively. They show similar patterns with $p=0.98$ in the main text.

\section{Fractions of residual couplings} 
\label{app:fraction_retained_ranks}
\begin{figure*}[!ht]
\centering
\includegraphics[width=0.97\textwidth]{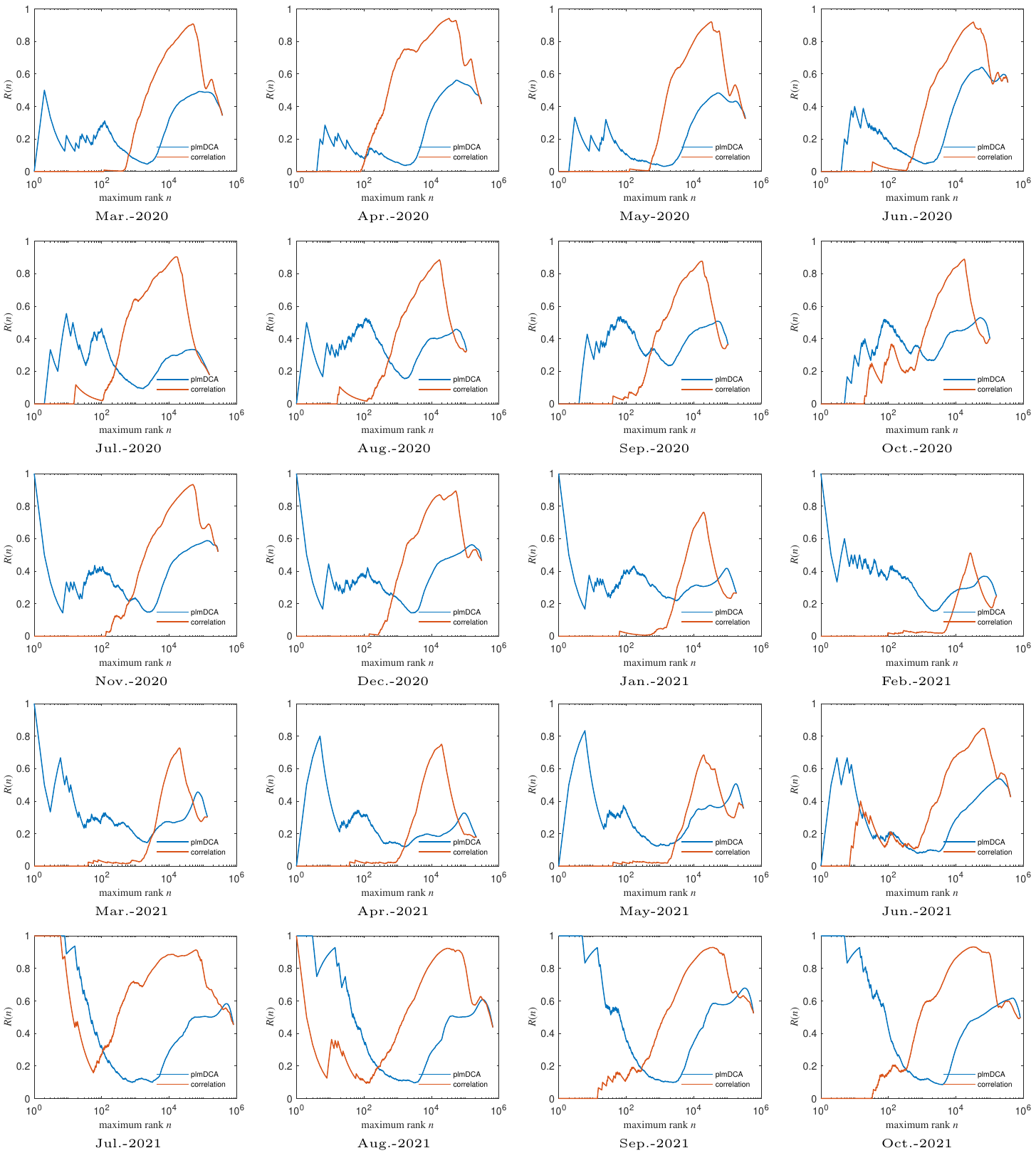}
\caption{Fraction of the residual epistatic couplings over the top-$k$ considered. Couplings between close sites ($<6$ bps), those with extrema in non-coding regions and those related to VoCs are excluded. Blue line for plmDCA while red for correlation analysis. }
\label{fig:Fraction_of_remain_ranks_all_months}
\end{figure*}

This appendix shows the fraction of residual (epistatic) couplings for plmDCA and correlation analysis as a function of the top-$k$ links considered, as shown in Fig. \ref{fig:Fraction_of_remain_ranks_all_months}. Data for the months Oct 2020 and Oct 2021 are also shown in Fig. \ref{fig:fraction_of_remain_links} of the main text.  For the highest ranks, plmDCA gives a greater fraction of true epistatic predictions with respect to correlation analysis. Couplings are removed if one or both $i$, $j$ meets/meet the removal conditions as described in Sec.(\ref{ss-FRC}).

\section{Circos plots with different filtering values for each month}

For each monthly dataset, three different $p$ values are employed for filtering the loci. If the percentage of a same major nucleotide along a column is larger than the given value (0.93, 0.95 and 0.98), then the column is discarded in the following DCA analysis. 

With plmDCA analysis, each pair of retained loci gets a score, which is related to the the epistasis between them. The pairwise epistatic links can be sorted by ranking the scores. Here, we plot the top-200 epistasis  with the ``Circos" software ~\cite{Krzywinski-2009a}.  Only those located in the coding region are shown. The short links \textit{i.e.}, those with distances between two terminals less than 4bps, are not included. Colored links are for those within top 50 ranks. Red ones for short links while blue ones for the long links. The greys are those within rank 51 to 200.

\begin{figure*}
\label{fig:2020_March_July}
\centering
\includegraphics[width=0.9\textwidth]{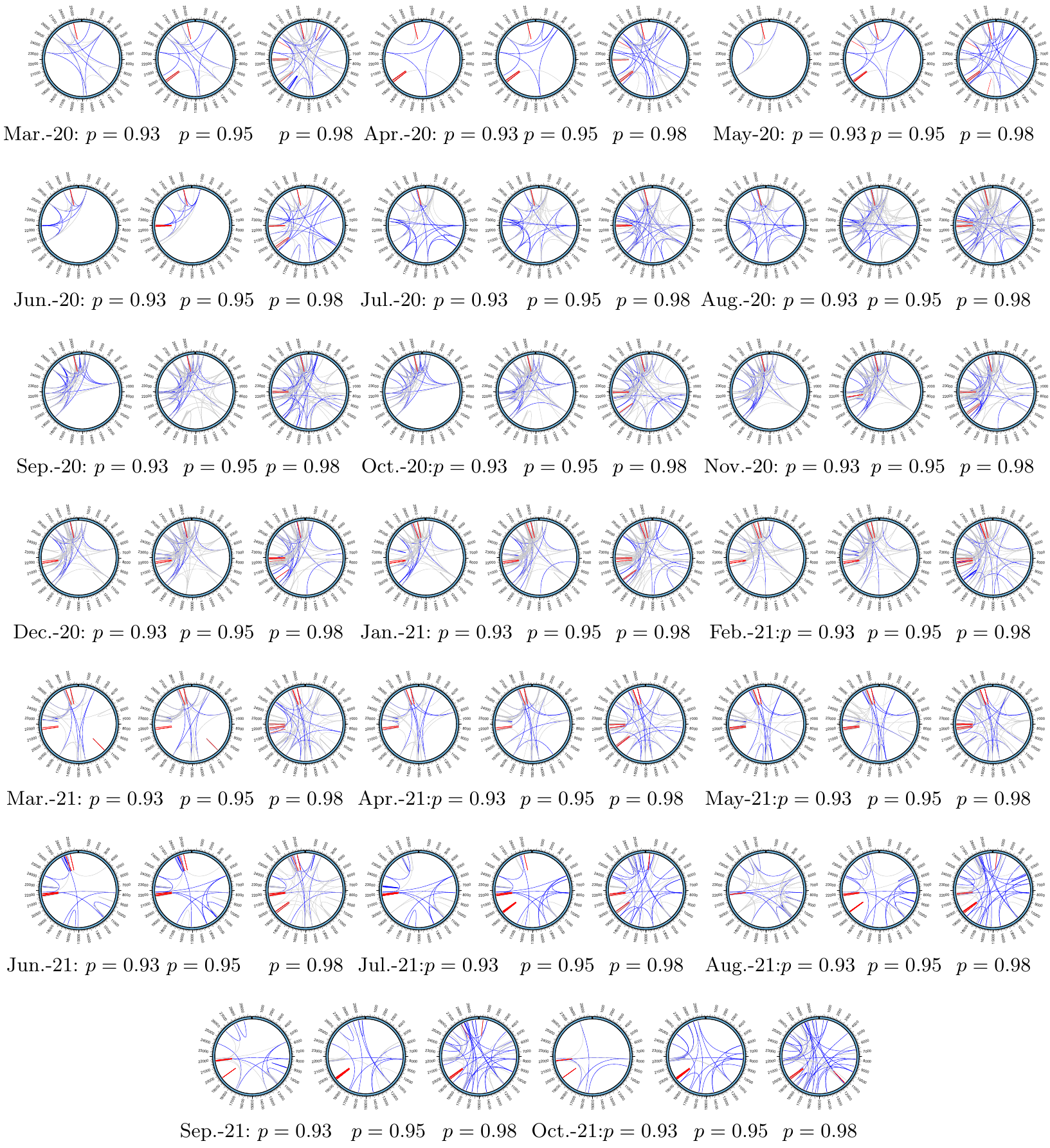}
\caption{Circos plots for epistatic links located in the coding region, top 200s ranks. The $p$ values for each month are selected as 93\%, 95\% and 98\% respectively.}
\end{figure*}

\section{Data resources}
\label{app:datasets}
All datasets listed below are available on github~\cite{Zeng-github}. 

\textbf{Dataset1}
    Accession\_IDs.xlsx\\
    The Accession IDs for the genomic sequences we used in the analysis. The prefix of each sequence "EPI\_ISL\_" is excluded to decrease the file size. This dataset is cut into two separate files further to satisfy the limitation of file size on Github, names as ``Dataset1-1-Mar-2020-May-2021-Accession\_IDs.xlsx" and ``Dataset1-2-Jun2021-Oct-2021-Accession\_IDs.xlsx" respectively on Github.

\textbf{Dataset2}
    p0.98\_plm\_Top200\_No\_3variants.xlsx\\
    This dataset contains selected links in top 200 plmDCA epistasic couplings, as ranked by their score. The plmDCA links shown in Fig. \ref{fig:ranks_plm_200_plmDCAs_vs_correlation}(a) and (b) in the main text are based on this dataset. Here, the links located in the non-coding region and with close locus ($\le 5$) and any loci included in alpha, beta, delta are excluded.
    
\textbf{Dataset3}
    p0.98\_Top\_2000\_CA\_No\_variants.xlsx\\  This dataset lists the sorted correlation scores which correspond to the dashed correlations in the middle and bottom Fig. \ref{fig:ranks_plm_200_plmDCAs_vs_correlation}(c) and (d). Similarly to its plm counterpart, links in coding region and the distance between loci is larger than 5bps are considered.  No variant is included.
    
\textbf{Dataset4}    
links\_with\_Spike\_locus\_or\_loci\_ranks.xlsx\\  The epistasis provided by plmDCA and correlation analysis are included in this dataset for each month. Only those within top-200s, for which the distance between two terminals is $> 5$ loci and whose both terminals located in the coding region are listed in the dataset. The genomic positions provided in Table 1 and 2 in the main text are based on this dataset.

\textbf{Dataset5} 
    protein\_aa\_mut\_for\_links\_in\_Dataset4.xlsx\\
    We provide the links within top 200s plmDCA scores that containing Spike terminals, for each month. The short links with loci located within 5 bps are discarded. Here, we also annotate the genes to which loci in the Dataset4 belong to and the corresponding amino acid mutations. The annotated genes and corresponding amino acid mutations in Table \ref{table:things-in-Spike-both} and \ref{table:things-in-Spike} in the main body of the manuscript are based on this dataset.

\nocite{*}

\bibliography{Covid19_arXiv}

\end{document}